\documentclass{article}
\usepackage[utf8]{inputenc}
\usepackage{fullpage}
\usepackage{enumitem}
\usepackage{booktabs}
\usepackage{multirow}
\usepackage{natbib}
\usepackage{subcaption}
\usepackage{graphicx}
\usepackage{amsmath}
\usepackage{amsfonts}
\usepackage{amssymb}
\usepackage{nicefrac}
\usepackage{xcolor}
\usepackage{hyperref}
\usepackage{csquotes}
\usepackage{mdframed}
\usepackage{tablefootnote}
\usepackage{authblk}
\usepackage{float}
\usepackage{placeins}
\usepackage{tabularx}
\usepackage{pifont}
\usepackage{listings}

\title{Who is a Better Matchmaker? Human vs. Algorithmic\\ Judge Assignment in a High-Stakes Startup Competition}
\author[1]{Sarina Xi}
\author[1]{Orelia Pi}
\author[2]{Miaomiao Zhang}
\author[2]{~\\Becca Xiong}
\author[2]{Jacqueline Ng Lane}
\author[1]{Nihar B. Shah}

\affil[ ]{\{sarinax, opi, nihars\}@cs.cmu.edu, \{mzhang, rxiong, jnlane\}@hbs.edu}
\affil[ ]{$~^{1}$Carnegie Mellon University, $~^{2}$Harvard University}

\date{}

\begin{document}
\maketitle

\begin{abstract}
There is growing interest in applying artificial intelligence (AI) to automate and support complex decision-making tasks. However, it remains unclear how algorithms compare to human judgment in contexts requiring semantic understanding and domain expertise. We examine this in the context of the judge assignment problem, matching submissions to suitably qualified judges. Specifically, we tackled this problem at the Harvard President’s Innovation Challenge, the university’s premier venture competition awarding over \$500,000 to student and alumni startups. This represents a real-world environment where high-quality judge assignment is essential. We developed an AI-based judge-assignment algorithm, Hybrid Lexical–Semantic Similarity Ensemble (HLSE), and deployed it at the competition. We then evaluated its performance against human expert assignments using blinded match-quality scores from judges on $309$ judge–venture pairs. Using a Mann–Whitney U statistic based test, we found no statistically significant difference in assignment quality between the two approaches ($AUC=0.48, p=0.40$); on average, algorithmic matches are rated $3.90$ and manual matches $3.94$ on a 5-point scale, where 5 indicates an excellent match. Furthermore, manual assignments that previously required a full week could be automated in several hours by the algorithm during deployment. These results demonstrate that HLSE achieves human-expert-level matching quality while offering greater scalability and efficiency, underscoring the potential of AI-driven solutions to support and enhance human decision-making for judge assignment in high-stakes settings.
\end{abstract}

\section{Introduction}
The rapid growth in innovative submissions requiring evaluations --- spanning academic conferences to startup competitions --- has given rise to a critical challenge: how to efficiently assign expert judges to submissions when submission volume is high and the individual judge evaluation bandwidth is limited. This assignment problem requires a sophisticated understanding of both judge expertise and submission content, complicated by the need to balance workload constraints \citep{criscuolo2017}, align domain expertise \citep{lane2024}, and ensure evaluation quality across diverse stakeholder needs \citep{boudreau2016}. 

In entrepreneurial competitions involving evaluations of startup ventures, individuals or teams present their business ideas to expert judges with the goal of securing funding, mentoring, or other support for their venture. Commonly organized by accelerators, corporations, universities, or government agencies, these competitions often span multiple tracks or categories, where each venture receives several reviews, and each judge reviews multiple ventures based on their expertise and backgrounds.

Recent advances in automating this assignment problem have shown promising results in academic peer review settings. Computer science conferences have been using algorithmic approaches to tackle the ``reviewer assignment'' problem for many years. In this setting, reviewer assignment is typically solved using a two-stage approach \citep[Section 3]{shah2021survey}: (1) computing similarity scores that estimate the alignment between each potential reviewer’s expertise and each submission, and (2) generating assignments based on these scores. The similarity computation process often uses publication histories or citation networks to get accurate similarity estimates \citep{cohan2020,ostendorff2022}. More recently, proposal reviewing agencies in astronomy have automated their reviewer assignment process~\citep{carpenter2025}. They use the history of proposal submissions to obtain similarity estimates.

However, a significant gap exists when applying these techniques to entrepreneurial competitions, and judge assignments herein are typically performed manually by a few program administrators who possess tacit knowledge of the context and process. The best performing algorithms for assigning reviewers to papers in peer review 
\citep{openreview_expertise}, such as SPECTER2 \citep{singh2023} and SciNCL \citep{ostendorff2022}, crucially rely on citation networks and past publications. However, judges in entrepreneurial competitions often lack such signals, making it much more difficult to algorithmically infer their domain knowledge. Moreover, entrepreneurial expertise spans both technical and business domains \citep{kacperczyk2017}, as evaluating market potential, business viability, and implementation strategy is essential for venture startups. Existing top-performing reviewer assignment algorithms are typically built on SciBERT \citep{beltagy2019}, a language model pre-trained on scientific corpora, which may not capture business-oriented expertise.

To address these challenges, we developed and evaluated a solution for the Harvard President's Innovation Challenge. Awarding more than \$500,000 annually, this is a high-stakes real-world setting where assignment quality directly impacts outcomes. We introduce HLSE (Hybrid Lexical–Semantic Similarity Ensemble), an ensemble similarity model that integrates three types of text representations to compute similarity estimates between judges and ventures: sparse TF-IDF vectors that contain many zeros and highlight distinctive keywords; dense transformer embeddings that contain mostly non-zero values and capture semantic meaning; and hybrid TF-IDF-weighted embeddings that merge both. Our approach follows the two-step pipeline used in academic peer review systems, using HLSE for similarity computation and PeerReview4All~\citep{stelmakh19} to assign matches according to the computed similarity scores.
Using this system, we investigate the key question: 
\begin{quote}
    \textit{Can algorithmic judge assignments match the quality of expert human judgment in such high-stakes entrepreneurial evaluation settings?}
\end{quote}

To answer this, we conducted the first direct empirical comparison between algorithmic and human expert assignments in entrepreneurial evaluation by collecting blinded self-reported match quality scores from judges.  Overall, our contributions are threefold:
\begin{itemize}
    \item We introduce HLSE, an ensemble approach that combines TF-IDF and transformer embeddings to calculate accurate similarity scores for the judge assignment problems. 
    \item We find that, across 309 judge-venture matches, human expert assignments achieved a mean match quality score of 3.94 compared to 3.90 for HLSE. A statistical test based on the Mann-Whitney U statistics showed no statistically significant difference between the two approaches based on these score ($AUC=0.48$, $p=0.40$), demonstrating human-level performance in a real-world deployment. 
    \item We demonstrate that HLSE significantly reduces assignment time from one week to several hours, offering a scalable solution to the growing demands of innovation assessment while maintaining quality standards.  

\end{itemize}
Although the datasets used for this work cannot be released due to data privacy agreements, our algorithm and evaluation methods can be found at: \url{https://github.com/xasayi/Automated-Judge-Assignment}.

\section{Related Work}

This work addresses the first stage of judge assignment: accurately computing similarity scores in the unique context of a large startup competition. In this section, we review existing approaches for both stages of judge assignment, examine related work evaluating similarity computation algorithms, and discuss the foundations behind our model, HLSE. 

\subsection{The Judge Assignment Problem}

\paragraph{Similarity score computation.} Numerous methods have been developed to determine the similarity between two descriptions, primarily using natural language processing and machine learning techniques. Traditional approaches such as TF-IDF and topic modeling \citep{ferilli2006, mimno07, anjum2019} work well on smaller datasets but often struggle to scale with large or evolving corpora.

To overcome these limitations, modern approaches increasingly utilize embedding-based methods. Sentences or documents are mapped into dense vector spaces using deep neural networks, capturing richer semantic relationships. For instance, SPECTER \citep{cohan2020} generates document embeddings using SciBERT \citep{beltagy2019} fine-tuned with citation links as a weak supervision signal. These citation-informed embeddings capture semantic similarity more effectively than purely text-based methods. Other works leverage citation graph structures to further inform document similarity \citep{ostendorff2022}. Such approaches provide richer contextual signals, particularly for documents with sparse or ambiguous text. Across all approaches, cosine similarity is commonly used for similarity computation, where a higher score indicates a more similar pair of embeddings.
\paragraph{Assignment algorithms.}
Once similarity scores have been computed, they are used to inform the assignment process. A common objective is to maximize the total similarity score across all assignments to enhance global match quality. This is used in the Toronto Paper Matching System \citep{Charlin2013TheTP}, which has been widely adopted by major Machine Learning conferences. However, maximizing global similarity is not always sufficient for fairness and work balance. Garg et al. \citep{garg2010} propose algorithms that balance judge preferences while ensuring equitable workloads and minimizing conflicts of interest. Similarly, Long et al. \citep{long2013good} frames the assignment process as a topic coverage problem such that each submission is reviewed by experts covering diverse relevant topics. Building on these foundations, Kobren et al. \citep{kobren2019} introduce a local fairness formulation that guarantees each submission receives a minimum threshold of judge expertise. To further mitigate manipulation and fraud risks, Jecmen et al. \citep{jecmen2020manipulation} proposes algorithms that use controlled randomness in the judge assignment process, with Xu et al. \citep{xu2024one} later extending these algorithms. 

In our work, we adopt the PeerReview4All algorithm \citep{stelmakh19}, which maximizes the minimum review quality assigned to any submission. By prioritizing the most disadvantaged submissions, this approach enhances fairness globally, particularly for submissions on niche or underrepresented topics, while maintaining overall assignment quality. The PeerReview4All algorithm has previously been used for assignment of reviewers to submissions in computer science conferences and astronomy proposal evaluations, resulting in a strong performance~\citep{carpenter2025,stelmakh19}.

\subsection{Evaluation of Similarity Computation Algorithms}

Evaluating the quality of similarity computations in judge assignment remains a core challenge. In the computer science peer review setting, many approaches use indirect estimates of the assignment quality, such as external expertise assessment \citep{mimno07, zhaoyue2022}. While informative, these methods can introduce noise due to limited context or incomplete information about the judges' backgrounds. To improve accuracy, other studies collect self-reported expertise ratings. For instance, Stelmakh et al. \citep{stelmakh2025} developed a gold-standard dataset of 477 blinded paper-reviewer pairs where reviewers report their expertise on papers. This benchmark has been used by OpenReview \citep{openreview_expertise}, a major computer science peer review platform used by several flagship conferences. While valuable, this benchmark is focused on academic peer review and may not generalize to domains like entrepreneurship, where expertise signals differ. Other studies, such as the one performed by Anjum et al. \citep{anjum2019}, gather self-reported expertise ratings from 33 reviewers after review completion to evaluate the quality of algorithmic assignments. 

Beyond computer science, fields such as astronomy have also started adopting and evaluating algorithmic judge assignment. Recently, Carpenter et al. \citep{carpenter2025} examined proposal review for the ALMA telescope, comparing mean similarity scores and self-reported reviewer expertise scores across multiple assignment cycles. Their findings show that automated assignment methods --- specifically, topic modeling to compute similarity scores, followed by PeerReview4All \citep{stelmakh19} --- increase both similarity scores and perceived reviewer expertise while significantly reducing manual effort. This provides important empirical evidence of the effective impact of ML-based algorithmic approaches outside of computer science peer review.

Given the lack of historical match-quality data in entrepreneurial settings, our experiment adopted a direct, empirical, and head-to-head comparison between algorithmic and manual assignments. Judges, blinded to assignment source, evaluated ventures matched independently by humans or the algorithm, then reported their perceived expertise fit. This allowed us to directly assess the alignment between algorithmic and human matches across 309 judge-venture pairs. To our knowledge, this is the first such empirical comparison beyond academic peer review, offering key insights into the effectiveness of algorithmic judge assignment in a high-stakes entrepreneurial context.

\subsection{Backbones of HLSE}
\textbf{TF-IDF methods.} Introduced in 1972 \citep{sparck1972}, TF-IDF remains a simple yet effective technique for text semantic similarity and has been widely used in a variety of applications, from text classification \citep{ liu2018, das2018} to information retrieval \citep{hiemstra2000, aizawa2003information, fautsch2010}. Term frequency (TF) measures how often a word appears in a document, while inverse document frequency (IDF) down-weights terms that appear in many documents and assigns higher weights to terms that occur in relatively few documents \citep[Section 6.2.1]{manning2009}. This property is particularly desirable in our application, where certain highly informative words may be frequent but confined to a small subset of documents; IDF ensures such terms still receive high weight.

In judge assignment, TF-IDF has gained widespread adoption through TPMS \citep{Charlin2013TheTP}, which has been integrated into major ML conferences. Different venues adapt TF-IDF differently based on their needs: some rely on titles and abstracts, others use full-text. Judge profiles incorporate prior assignments and the specific TF-IDF method used can be tuned \citep{openreview_expertise}. Despite advances in deep learning, recent studies suggest TF-IDF remains competitive \citep{gonzálezmárquez2024}. Stelmakh et al. \citep{stelmakh2025} finds that TPMS with full-text inputs matches or outperforms state-of-the-art embedding models and large language models in assignment accuracy.

\paragraph{Transformer-based embedding methods.}
Embedding methods use pretrained language models to transform documents into dense vector representations. Due to the input length limitations of transformers, these methods typically rely on titles and abstracts rather than full-text \citep{cohan2020}. Prior research shows embeddings perform better on short texts, whereas TF-IDF excels on longer texts \citep{meijer2021documentembeddingscientificarticles}. A common method used in judge assignment for computer science peer review is SPECTER \citep{cohan2020}, which uses the title and abstract of papers. It is trained with a citation-informed contrastive loss, using cited papers as positive examples and non-cited work as negative examples for semantic similarity \citep{cohan2020}. Building on this framework, SciNCL \citep{ostendorff2022} uses harder examples to improve embedding quality, while SPECTER2 \citep{singh2023} extends the training corpus and incorporates multi-task learning. All of these models use SciBERT \citep{beltagy2019}, a variant of BERT pre-trained in a large scientific corpus. However, these models remain domain-specific to scientific language and may underperform in entrepreneurial contexts, which have different vocabularies and semantic nuances.

\paragraph{Hybrid similarity methods.} While hybrid approaches combining TF-IDF and embeddings have been explored in NLP tasks, none are used in judge assignment. For example, Agarwal et al. \citep{agarwal2019} and De Boom et al. \citep{deboom2016} compute an informed sum of TF-IDF weighted word embeddings to improve author clustering and representation learning while Didi et al. \citep{didi2022covid} uses a TF-IDF weighted sum of word embeddings for classification of COVID-19 tweets. Other works have explored alternative weighting schemes. Arora et al. \citep{arora2017} use a weighted mean of word embeddings with Smoothed Inverse Frequency (SIF), defined as $a/(a+f_w)$, where $a$ is a parameter and $f_w$ is the frequency of word $w$ in the corpus. While they demonstrate that SIF-based embeddings outperform neural methods such as RNNs and LSTMs on several tasks, this weighting can undervalue informative terms in our dataset that occur frequently but only within a small number of documents. All of these prior methods rely on static embeddings such as GloVe \citep{penningtno2014}, which cannot capture context-dependent meanings. In contrast, we use transformer-based embeddings, which model semantic nuances and can distinguish between different meanings of the same word. Consistent with this choice, Joshi et al. \citep{joshi2020} report that combining TF-IDF with transformer-based embeddings can yield up to a 36\% relative improvement on fine-grained tasks.

Beyond using hybrid models, HLSE employs an ensemble that integrates similarity signals from TF-IDF, transformer-based, and hybrid representations. By aggregating these signals, the ensemble is designed to reduce reliance on the strengths or weaknesses of any single representation. This strategy has been shown to often have better performance than individual models \citep{Opitz_1999, diet2000}.

\section{Problem Setup and Data}
This section describes the problem setup, including the real-world deployment context of our model, the problem statement, the challenges of manual judge assignment, and an overview of the training data used for model development.

\paragraph{Real-world context.} This work is set in the context of the Harvard President’s Innovation Challenge, which annually attracts hundreds of early-stage student and alumni ventures from diverse sectors including healthcare, fintech, consumer products, and more. These ventures span various stages and compete for more than \$500,000 in funding and support. All ventures undergo screening by an internal selection committee, where semifinalists are then evaluated by external judges. This external judging pool consists of high-profile judges from various institutional, functional and domain backgrounds, including experienced investors, established entrepreneurs, and senior professionals with deep technical or operational expertise. Many have led successful business exits or held senior leadership or scientist roles in top-tier firms or academic research institutions. To ensure impartiality and consistency in venture evaluation, judges are assessed for relevance to ventures and provided with structured rubric criteria to standardize evaluations.

\paragraph{Problem statement.} For the 2025 competition, 231 judges and 101 semifinalist ventures were considered for automated assignment. The core task is to assign each semifinalist venture to a panel of qualified judges whose expertise, background, and experience align with the venture’s industry, technology, and market focus. At the same time, the assignment must satisfy logistical and fairness constraints, including:
\begin{itemize}
    \item Load balancing: Each judge can evaluate at most 7 ventures and each venture must have exactly 12 judges assigned to it.
    \item Track constraints: Judges are matched only to ventures in the same tracks (``Open", ``Social Impact", ``Healthcare and Life Sciences").
    \item Conflict of interest exclusions: Assignments where a judge has personal, professional, or financial ties to a venture must be avoided.
\end{itemize}

These constraints align with competition goals and ensure standardization compared to previous practice (see Table~\ref{table:historical_statistics}).

\begin{table}[htbp]
\centering
\begin{tabular}{|c|c|c|c|c|c|c|}
\hline
\textbf{Year} & \multicolumn{3}{c|}{\textbf{Ventures per Judge}} & \multicolumn{3}{c|}{\textbf{Judges per Venture}} \\
\cline{2-7}
 & Min & Mean & Max & Min & Mean & Max \\
\hline
2021 & 2 & 7 & 12 & 8 & 13 & 22 \\
2022 & 2 & 7 & 12 & 8 & 12 & 17 \\
2023 & 1 & 8 & 10 & 7 & 16 & 27 \\
2024 & 2 & 8 & 13 & 9 & 16 & 23 \\
\hline
\end{tabular}
\caption{Historical venture-judge assignments summary statistics.}
\label{table:historical_statistics}
\end{table}

\paragraph{Limitations of manual process.} From 2021 to 2024, the process of assessing and assigning judges to semifinalist ventures was performed manually by an internal program administrator using informal heuristics. While this approach benefited from continual assignment experience and institutional knowledge, it became increasingly unsustainable due to the several reasons:
\begin{itemize}
    \item \emph{Growing scale}: The number of ventures grew from 112 to 133 and judges from 201 to 281. This expanded the potential matching space by 66\%, from 22,512 (112$\times$201) to 37,373 (133$\times$281).
    \item \emph{Staff turnover}: Knowledge of past judge–venture matching decisions was difficult to retain and transfer between program administrators.
    \item \emph{Heuristic-based groupings}: Manual groupings used by the program administrator prioritized perceived quality within small clusters rather than global semantic alignment. Prior work has shown that grouping-based constraints, whether for diversity~\citep{Benabbou_2020}, two-phase review design~\citep{jecmen2021}, or strategy proofing in peer review~\citep{dhull2022}, can reduce overall match quality compared to individualized matching.
\end{itemize}
Manually constructing such an assignment on this scale risks inferior match quality and fairness concerns. These challenges, combined with the need for scalable and high-quality judge-venture matches, highlight the need for a computationally-driven solution.

\paragraph{Training data.} To train our assignment model, we received four years (2021–2024) of historical competition data from the organizers, consisting of three components:

\begin{itemize}
    \item Venture applications, encompassing pitch descriptions, track, venture industry, venture problem description, and solution details. 
    \item Judge profiles, containing short bios, preferred track, areas of expertise, and industries interests. If the profile length contained fewer than 50 words, we supplemented it with publicly available data collected and verified by the competition organizers, sourced from their LinkedIn profiles and company websites.
    \item Historical assignments, consisting of 1,400 to 2,100 judge-venture matches each year. 
\end{itemize}
While these records could serve as training data --- using venture and judge text descriptions as inputs and past assignments as labels --- the informal manual heuristics underlying these past matches make them unreliable as ground truth. Moreover, the historical assignments provide only a binary signal indicating whether a judge-venture pair was assigned or not, with no measure of match quality. To create a more reliable and informative training dataset, we manually scored a subset of 105 judge-venture pairs for match quality on a 1-5 scale, where 1 denotes no relevant judge expertise alignment, and 5 reflects precise alignment of judge expertise with the venture's area. These scores were assigned without relying on any pre-existing groupings by members of the research team as well as the competition staff, both of whom have deep knowledge of the competition context and assignment process. This curated dataset served as a trusted ground-truth dataset for match quality to develop our model.

In addition, to prepare the input judge profile and venture application text description for similarity computations, we only retained a subset of relevant fields from the raw competition data (see Appendix \ref{appendix:feature_selection} for details). Then, we sanitized the input text using established methods \citep{Charlin2013TheTP, xu2020}. 

\section{HLSE Method}
The HLSE method is an ensemble model that integrates multiple similarity computation models as base learners. These base learners include TF-IDF models, transformer-based embedding models, and hybrid models that combine IDF weighting with transformer-based representations. In this section, we first describe two hybrid models used in HLSE: one based on document-level similarity and the other on token-level similarity. We then outline the specific base models used. Finally, we explain how the outputs of all base learners are combined into a unified similarity ensemble.

\newcommand{\tokenEmbedding}{T}
\newcommand{\idfWeight}{w}
\newcommand{\seqLen}{L}
\newcommand{\embeddingDim}{d}
\newcommand{\seqLenIndex}{\mathbb{\ell}}
\newcommand{\docEmbedding}{D}
\newcommand{\simScore}{s}
\newcommand{\simMat}{S}
\newcommand{\indexi}{i}
\newcommand{\indexj}{j}

\paragraph{Document level mean hybrid similarity.}
To formalize the hybrid approaches, we introduce two pieces of notation. Let $\tokenEmbedding \in \mathbb{R}^{\embeddingDim \times \seqLen}$ denote the transformer-based token embeddings with a sequence length $\seqLen$ and embedding dimension $\embeddingDim$. Further, let $\idfWeight \in \mathbb{R}^{\seqLen}_{> 0}$ denote IDF weights corresponding to each token. Using this notation, we compute the document embedding, $\docEmbedding$, for each input (venture application or judge profile) as: 
$\docEmbedding = \frac{\tokenEmbedding\cdot \idfWeight}{\|\idfWeight\|_1}$

Given such embeddings for any two documents $\docEmbedding^{(1)}$ and $\docEmbedding^{(2)}$, we use cosine similarity to evaluate the semantic similarity between them: $\frac{\docEmbedding^{(1)} \cdot \docEmbedding^{(2)}}{\|\docEmbedding^{(1)}\|_2 \cdot \|\docEmbedding^{(2)}\|_2} \in \mathbb{R}$. This yields a value in $[-1, 1]$, where $1$ denotes maximum similarity and $-1$ denotes maximum divergence. 

\paragraph{Token level mean hybrid similarity.} While the document-level method aggregates embeddings before similarity calculation, the token-level approach calculates similarity at the token stage and then applies the IDF. This better compares how closely individual tokens from different documents align. For two token embedding columns $\tokenEmbedding_\indexi^{(1)}$ at the $i^{\text{th}}$ column in the first document and $\tokenEmbedding_\indexj^{(2)}$ at the $j^{\text{th}}$ column in the second document, we calculate the IDF-weighted cosine similarity between them 
$\simScore_{\indexi,\indexj
}
=\frac{\tokenEmbedding_\indexi^{(1)} \cdot \tokenEmbedding_\indexj^{(2)}}{\|\tokenEmbedding_\indexi^{(1)}\|_2 \cdot \|\tokenEmbedding_\indexj^{(2)}\|_2}\cdot w_i^{(1)}w_j^{(2)} \in \mathbb{R}$.
We then compute the overall similarity over the set $\mathcal{\simMat}$ of all token level similarities: $\frac{1}{|\mathcal{\simMat}|} \sum_{\simScore_{\indexi, \indexj} \in \mathcal{\simMat}} \simScore_{\indexi, \indexj} \in \mathbb{R}$. This yields a similarity score where a larger positive value indicates that the two documents are more similar.

\paragraph{Specific base learners.} The HLSE model consists of three types of base learners: TF-IDF models, transformer-based embedding models, and hybrid models. For TF-IDF, we implement two main TF-IDF approaches, each trained with and without an expanded Wikipedia corpus to enhance IDF generalization, resulting in 4 variants in total. The first approach is a standard TF-IDF model with IDF smoothing. The second is an augmented TF-IDF model by~\citep{xu2020}, which mitigates bias toward longer documents through an adjusted the TF formulation and using a non-smoothed IDF. The supplementary Wikipedia corpus, retrieved via WikipediaAPI, contains 2,108 documents with a mean length of 250 words and standard deviation of 29 words. 

For embedding models, we use 4 transformer models. These include general-purpose models, such as BERT, and more specialized sentence transformers like RoBERTa and MPNet. We also selected the top performing model on the MTEB \citep{mteb2023} Semantic Textual Similarity (STS) benchmark, prioritizing resource-efficient architectures due to limited local compute. Notably, \textit{GIST-Embedding-v0} is a lightweight model with a sub-1~GB memory footprint and no multilingual tuning, with strong performance on STS tasks. For each of the embedding models, we derive 2 base learners by computing similarity at either the document level or the token level. For hybrid models, we combine the two hybrid similarity computation schemes detailed earlier with four IDF schemes, yielding 8 variants for each embedding model. In addition to these models, we also included GPT-4o using zero, one, and few shot prompting as detailed in Appendix \ref{appendix:llm}. In total, we have 4 TF-IDF base learners, 8 embedding base learners, 32 hybrid base learners, and 3 GPT-4o base learners.

\paragraph{Ensemble learning.}\label{section:ensemble}
After computing similarity scores from the base models, we combine them into a single ensemble similarity score. Our goal is to learn nonnegative weights such that the weighted combination of base similarity scores best approximates the ground truth match quality scores. We achieve this by solving a linear regression problem with convexity constraints, requiring the nonnegative weights to sum to one. The ensemble weights are learned by minimizing the mean squared error between the ground-truth labels and the weighted predictions. We perform five-fold cross-validation, and the final ensemble prediction is obtained as a weighted sum of the base model similarity scores using the learned weights. Only 5 of 47 base learners received weights greater than 0.01 in every fold. These five models were retained in the final ensemble, with the remaining dropped. The specific models that make up the ensemble are shown in Table \ref{table:base_learners}. On the match quality training dataset, 85\% of the ensemble’s predicted scores fall within one point of the ground-truth, and 37\% match the ground truth exactly.

\begin{table}[h]
\centering
\begin{tabular}{|l|l|}
\hline
\textbf{Base Learner Type} & \textbf{Specific method}\\
\hline
TF-IDF & TFIDF with Wikipedia Weighted IDF\\
\hline
Embedding & BERT Token Mean Pooled \\
Embedding & MPNet Document Mean Pooled \\
Embedding & RoBERTa Document Mean Pooled \\
\hline
Hybrid & BERT IDF Weighted Token Pool \\

\hline
\end{tabular}
\caption{The base models in the ensemble learner.}
\label{table:base_learners}
\end{table}

\section{Algorithmic versus Manual Assignment Experiment} \label{section: survey}

\begin{figure}[htbp]
    \centering
    \includegraphics[width=0.6\linewidth]{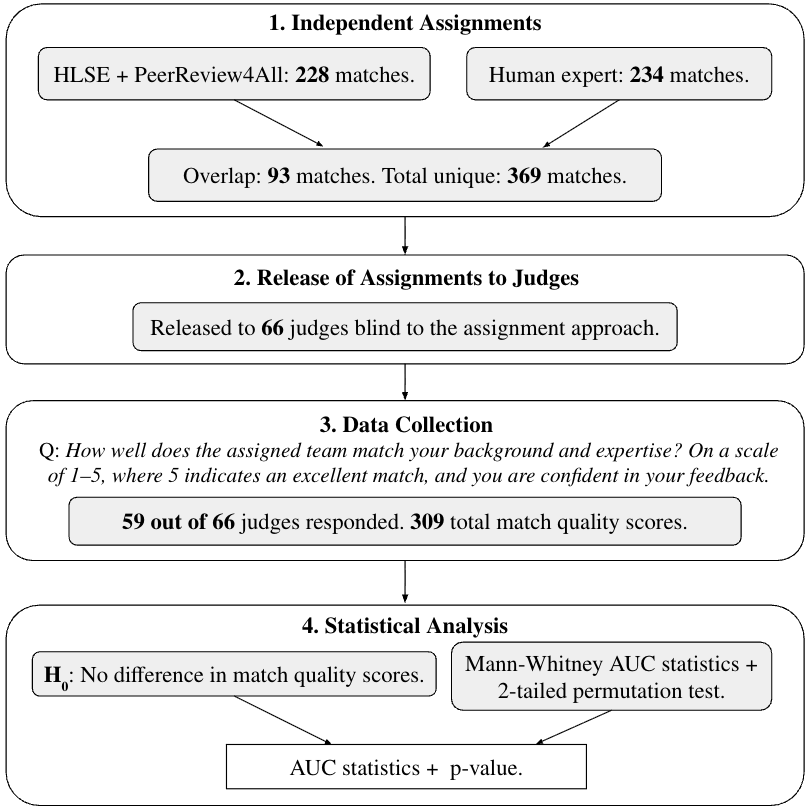}
    \caption{Workflow for algorithmic versus manual assignment experiment.}
    \label{figure:workflow}
\end{figure}

To compare the quality of algorithmic and manual judge assignments, we conducted an experiment based on the judge's self-reported assessment of expertise alignment with the ventures they were matched with. Figure\ref{figure:workflow} outlines the experimental workflow. 

\paragraph{Experiment design.} To isolate the effect of the assignment method, manual and algorithmic assignments were performed independently. Per the competition organizers' request, the manual assignment was only performed for ventures in one of the tracks. This track was the smallest in size, consisting of 19 ventures and 66 judges. Each venture in this smallest track received 11 to 13 manual judge assignments and exactly 12 algorithmic assignments. This resulted in 369 unique judge-venture pairs, with 228 algorithmic matches, 234 manual matches, and 93 overlapping matches. All such algorithmic matches were accepted by the organizers without adjustment. In addition, judges were blind to the assignment method. For the other two tracks, which used algorithmic assignment but were not a part of this experiment, the competition organizers performed manual refinement on the algorithmic assignments before releasing them to the judges.

\paragraph{Outcome measurement.} Judges were asked the following question with radio buttons ranging from 1 to 5 as part of their evaluation form: 

\begin{quote}
\textit{``How well does the assigned team match your background and expertise? On a scale of 1–5, where 5 indicates an excellent match, and you are confident in your feedback."}
\end{quote}

\noindent To ensure consistency with other questions asked in the evaluation form unrelated to this experiment, the order of the options is not randomized. Of the 66 judges who indicated their preference to participate in reviewing venture applications from this track, 59 provided responses, yielding 309 total match quality scores.

\paragraph{Statistical test.} To compare the two methods, we test the null hypothesis:
\begin{quote}
\emph{There is no systematic difference in match quality between algorithmic and manual assignments.}
\end{quote}

We use a test based on the Mann–Whitney U statistic to assess whether there is a statistically significant difference in match quality between the two assignment methods. This nonparametric test is well-suited for comparing ordinal data from independent groups. Our setup considers each venture as a unit of comparison, with judges either assigned manually or via the algorithm. To find the statistical significance, we performed a two-tailed permutation test. The specific procedure is detailed in Appendix. \ref{appendix:statistical_test}.

\section{Results} 
We deployed HLSE to compute the similarity scores between judges and ventures in the 2025 competition. In this section, we analyze the results from the experimental procedure, the time taken for manual versus algorithmic assignment, and an internal post-deployment audit.

\paragraph{Experiment results.} Following the experimental procedure in Section ~\ref{section: survey}, we collected self-reported match quality scores for 309 judge-venture pairs. Table ~\ref{table:match_quality_summary} summarizes the comparison between the manual and algorithmic assignments. The mean self-reported match quality scores are comparable for both approaches: 3.94 for manually assigned matches and 3.90 for algorithmically assigned matches. The statistical test yields an AUC statistic of {0.48} and a p-value of {0.40}. Here, an AUC of 0.5 indicates no difference in score distribution between the two groups, while an AUC closer to 0 or 1 would indicate that one group always has higher scores than the other. These results indicate no statistically significant difference in match quality between the two assignment approaches, thus does not provide evidence for rejecting the null hypothesis. In addition, we observe that there are 82 overlapping matches where judge-venture pairs that have been assigned by both HLSE and the human expert. These matches have a mean score of 4.10.

\begin{table}[h]
\centering
\begin{tabular}{|l|c|c|}
\hline
 & \textbf{Manual} & \textbf{Algorithmic} \\
\hline
Number of scores & 197 & 194 \\
Mean quality score & 3.94 ($\pm$0.93) & 3.90 ($\pm$0.94) \\
Number of overlapping matches & \multicolumn{2}{c|}{82 (included in both)} \\
Mean score for overlapping matches & \multicolumn{2}{c|}{4.10 ($\pm$0.94)} \\
\hline
\end{tabular}
\caption{Summary mean and standard deviation of the match quality scores from manual versus algorithmic assignments.}
\label{table:match_quality_summary}
\end{table}

\paragraph{Manual versus algorithmic assignment time.} We compared the time required for algorithmic versus manual assignments. Manual assignment times were provided by the competition organizers, while algorithmic times included both the actual runtime of the algorithm (on an Apple M1 chip with 16GB of RAM) and the time it took to scrape publicly available supplementary information for judges. This data collection process was performed using automated tools by the competition organizers and is necessary each year because the information can change over time. The time required to write the scrapers and develop the algorithm is not included since it is an amortized cost, and the developed code can be reused in the future. 

\begin{figure}[h]
    \centering
    \includegraphics[width=0.7\linewidth]{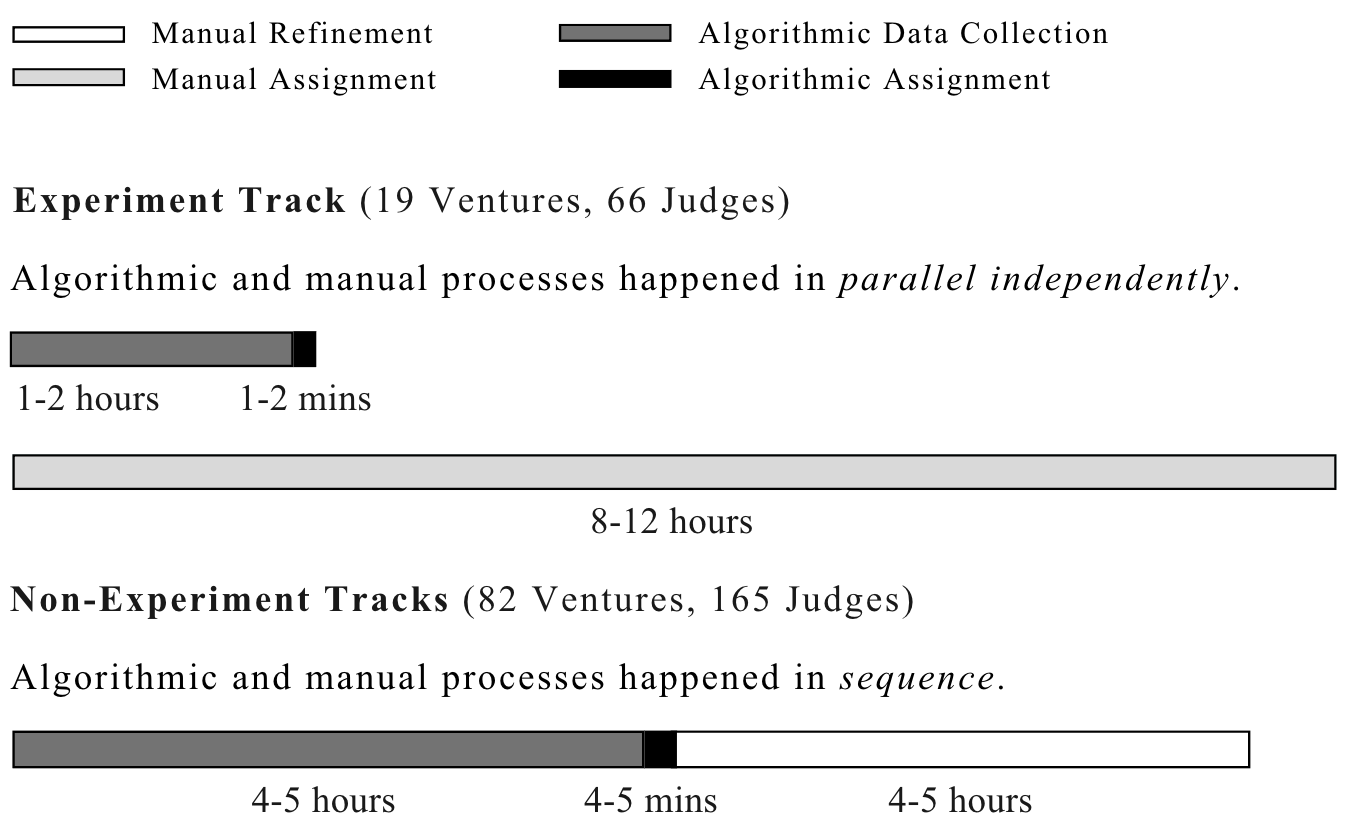}
    \caption{Overview for time taken for 2025 assignments.}
    \label{figure:time_comparison}
\end{figure}

For 2021 to 2024, the manual assignment process took about \textit{a full week} each year for around 112-133 ventures and 201-281 judges. This included reading judge profiles and venture applications, searching online for additional information about judges’ expertise, and manually matching judges to ventures. Figure \ref{figure:time_comparison} summarizes the time spent on these tasks for 101 ventures and 231 judges in 2025, split between the experiment and non-experiment tracks.

In 2025, manual and algorithmic assignments were performed independently on the experiment track. The manual assignment process took 8–12 hours, while the algorithmic data collection and assignment took 1–2 hours and 1–2 minutes, respectively. For the non-experiment tracks, algorithmic data collection took 4–5 hours and the algorithmic assignment took 4–5 minutes, followed by an additional 4–5 hours of manual refinement.

Although the experiment track was much smaller in scale, its manual assignment alone took roughly the same amount of time as the combined manual refinement and algorithmic assignment across all non-experiment tracks. Moreover, since our experiment results show that the algorithmic and manual approaches achieved comparable match quality, this indicates that HLSE can generate reliable similarity scores that maintain assignment quality even without refinement. Compared to the week-long process in previous years, the 2025 algorithmic approach reduced the total assignment time to approximately 5–7 hours.

\paragraph{Internal audit.} Post-deployment, we conducted an internal audit in which we discovered a minor implementation flaw affecting all hybrid models due to a tokenization mismatch. While this flaw introduced some noise in the models’ similarity estimates, it did not meaningfully change the ensemble’s overall behavior. We recalculated the AUC statistic, obtaining 0.49 with a p-value of 0.76, indicating no statistically significant difference between algorithmic and manual assignment outcomes. Comparing the old against the new algorithmic matches, we find that there are 168 overlapping matches, with a mean expert-assigned quality score of 3.92 (standard deviation = 0.95). The 26 matches assigned only by the old ensemble had a mean score of 3.77 (standard deviation = 0.80), while the 5 matches assigned only by the new ensemble had a mean score of 4.4 (standard deviation = 0.80). While there is some difference in the assigned judge-venture pairs, removing the flawed hybrid model did not produce a statistically significant change in the distribution of assignment quality scores, and the mean expert-assigned scores remain comparable between the old and new ensembles. Unintentionally, these results demonstrate that our algorithm is robust against some noise in the base learners.

\section{Conclusions}
Assigning judges with the appropriate expertise to submissions is a critical challenge in various domains that require high-quality evaluation. Existing similarity algorithms that are deployed for reviewer assignment in scientific peer review are not directly applicable to other domains. In this work, we tackle judge assignment in the context of a competitive entrepreneurial challenge. We developed an ensemble similarity model, HLSE, and deployed it in the Harvard President's Innovation Challenge, a venture competition that awards \$500,000 in prizes annually. Our results demonstrate that algorithmic judge assignment performs on par with expert human judgment in this entrepreneurial context. Matches produced by HLSE are statistically indistinguishable from matches produced by human experts based on self-reported match quality scores. Furthermore, the algorithmic approach is significantly more scalable and time efficient, reducing the assignment time from a full week to hours. Below, we discuss key lessons from HLSE's deployment at the 2025 competition and outline directions for future work.

\paragraph{Curating ground truth datasets.} Early in development, we observed that the historical assignment data were often inconsistent or suboptimal. This reflects a common challenge in real-world settings, where the data available can be noisy, incomplete, or collected under varying standards. Such issues can significantly hinder meaningful evaluation. To address this, we curated a high-quality, expert-informed match quality dataset using insight and expertise from the organizers, providing a more reliable and consistent ground truth dataset to develop our method. This experience highlights the importance of developing reliable ground truth datasets that reflect real-world expectation, especially when existing data is unreliable or unavailable.

\paragraph{Empirical stability of the ensemble.} A post-deployment audit uncovered a minor implementation flaw in the tokenization scheme for hybrid models, introducing some noise to the similarity estimates. Despite this, the ensemble's overall performance remained stable, and most individual matches remained consistent between the old and revised ensembles. These results indicate that, in this deployment, the ensemble performance was largely unaffected by the flaws in a subset of base models, (unintentionally) providing empirical evidence of system stability in practice.

\paragraph{Human-AI collaboration insight.} Our experiment showed that algorithmic assignments using HLSE for computing similarity scores and PeerReview4All \citep{stelmakh19} for assignment performed on par with manual expert assignments. Importantly, even with human refinement for the non-experiment tracks, the combined assignment process remains significantly faster than full manual assignment. This suggests that AI-assisted workflows can effectively balance scalability and efficiency with human oversight, offering a practical path to high-quality assignments.

\paragraph{Future work. } Building on the lessons learned, we propose several directions to address current limitations and advance this work:
\begin{itemize}

\item \emph{Expanded cross-domain and longitudinal validation:} To evaluate the generalization of HLSE, future work can focus on extending the algorithmic versus human evaluation across a larger sample size, more competition tracks, different types of review systems (e.g., hackathons), and at other institutions. Deploying HLSE and PeerReview4All~\citep{stelmakh19} over multiple annual cycles and varying institutional contexts would allow better assessment of the stability of the method under different data regimes, judge pools, and domain characteristics. Finally, this can also help isolate HLSE's performance from contextual effects tied to any single deployment. 

\item \emph{Prevent gaming of automated assignment methods:} Recent work~\citep{eisenhofer2023no, jhihyi2024} has shown that automated assignment algorithms are susceptible to manipulation, whereby strategic text modifications can be used to exploit the method and obtain favorable matches. Such vulnerabilities can facilitate fraudulent activities, including collusion rings~\citep{littman2021collusion, jecmen2020manipulation} and identity theft~\citep{shah2025identity}. This poses a serious risk to the integrity of such automated methods, especially in high-stakes settings such as venture competitions. Future work can investigate how to protect against these adverse mechanisms to protect HLSE against such adversarial behavior, building on the ideas from~\cite{jhihyi2024,jecmen2020manipulation,shah2025identity} that are widely deployed in computer science conference reviewing.

\item \emph{Human-in-the-loop feedback:} Building on our current results, future work could investigate how integrating human input into algorithmic assignments affects match quality and efficiency at scale. For instance, one could explore human-refined algorithmic assignments compared with fully manual or purely algorithmic assignments to better understand the specific benefits of human-AI collaboration.

\end{itemize}

\section*{Acknowledgments}
This work was done under the approval of the Carnegie Mellon University Institutional Review Board. We gratefully acknowledge the contributions and insights of the organizing team and collaborators at Harvard Innovation Labs. This work was supported in part by NSF 1942124 and ONR N000142512346.

\bibliography{bibtex}
\bibliographystyle{alpha}

~\\~\\~\\
\begin{appendix}
\noindent{\Large \textbf{Appendices}}

\section{Data Subset}\label{appendix:feature_selection}
The historical dataset used in this work was highly heterogeneous, comprising 88 columns: 43 describing ventures and 45 describing judges. These columns varied substantially in format and quality, ranging from detailed paragraphs (e.g., venture descriptions and judge bios) to sparsely populated fields containing single-word entries (e.g., industry). We conducted a manual inspection to remove redundant or low-utility fields such as internal identifiers, timestamps, or sparse columns with mostly empty entries. This reduces the risk of truncating critical information when working with transformer input length limitations, and removes potential noise from irrelevant fields. As a result, we selected:

\begin{itemize}
    \item 9 columns for ventures, including key fields such as industry, description, pitch, problem, and solution.
    \item 35 columns for judges, encompassing short biography, areas of expertise, and industry focus.
\end{itemize}

In 2025, the competition organizers modified the data schema, resulting in a new column structure. To ensure consistency across years, we mapped the updated fields to their historical counterparts and obtained 8 venture columns and 15 judge columns that best represented the features identified in the earlier years. 

\section{Using GPT-4o as a Judge}\label{appendix:llm}
We queried \textit{gpt-4o-2024-08-06} with a temperature of 0 (no randomness) to produce similarity scores given a venture and judge input. We used zero, one, and few shot learning with the following prompt where the exact examples provided in the prompt are omitted:

\begin{lstlisting}
    You are a manual assigner who is responsible for assigning judges to ventures for a venture competition. You are given two blocks of text, D1 being the venture and D2 being the judge. You need to assess whether the judge is a good match to the venture considering whether the judge has technical expertise in the venture\'s area. Rate the match on a scale of 1-5 (5 is good, 1 is bad). Internalize your reasoning and only output a number between 1 to 5.
    D1: <description of the venture>
    D2: <description of the judge>
\end{lstlisting}

We compare the scores generated by GPT-4o against the ground truth match quality scores and find the Kendall Tau-b rank correlation, $\tau$, as well as the associated p-value, $p$, between them. The null hypothesis for the p-value is that the generated and ground-truth scores are independent. For the zero-shot setting, we obtain $\tau = -0.02$ ($p = 0.822$); for one-shot, $\tau = 0.00$ ($p = 0.980$); and for few-shot (two examples), $\tau = 0.02$ ($p = 0.752$).

\section{Statistical Test}
\label{appendix:statistical_test}
Here we provide details about the statistical test for the algorithmic versus manual experiment. 
\subsection{Mann-Whitney U Statistic}

\newcommand{\numVen}{k}
\newcommand{\numAutoi}{n_{i}^A}
\newcommand{\numManui}{n_{i}^M}
\newcommand{\numOveri}{n_{i}^O}
\newcommand{\auto}{A}
\newcommand{\manu}{M}
\newcommand{\overlap}{O}
\newcommand{\autoIndex}{a}
\newcommand{\manuIndex}{m}
\newcommand{\overlapIndex}{o}
\newcommand{\ustat}{U}
\newcommand{\weighting}{w}
\newcommand{\aucNorm}{V}
\newcommand{\pval}{p}

We compare assignment quality between algorithmic and manual assignments using an adjusted Mann–Whitney \(U\) statistic. Since some assignments appear in both groups, we adjust the statistic to avoid double counting the overlaps. For each venture \(i \in \{1,\dots,\numVen\}\), where $\numVen$ is the total number of ventures, let
\begin{itemize}
    \item \(\auto_i\): set of algorithm-only assignment scores, \(\numAutoi=|\auto_i|\),
    \item \(\manu_i\): set of manual-only assignment scores, \(\numManui = |\manu_i|\),
    \item \(\overlap_i\): set of overlapping assignment scores, \( \numOveri = |\overlap_i|\).
\end{itemize}
where the scores are in \(\{1,\dots,5\}\). 

The Mann-Whitney score function between two scores \(x,y\) is the following, where \(\mathbb{I}[\cdot]\) is the indicator function:
\[
h(x,y) = \mathbb{I}[x > y] + \tfrac{1}{2}\mathbb{I}[x=y].
\]

Then, the per-venture U statistic is
\[
\ustat_i = \sum_{\autoIndex \in \auto_i \cup \overlap_i} ~~\sum_{\manuIndex \in \manu_i \cup \overlap_i} h(\autoIndex,\manuIndex).
\]

Since the overlapping assignments contribute equally to both groups and can artificially inflate the U statistics, we subtract the self-comparison term:

\[
\ustat_i^\overlapIndex = \sum_{\overlapIndex_1 \in \overlap_i} ~~\sum_{\overlapIndex_2 \in \overlap_i} h(\overlapIndex_1,\overlapIndex_2).
\]

The overlap-adjusted per venture AUC statistic is:
\[
T_i = \frac{\ustat_i - \ustat_i^\overlapIndex}{\aucNorm_i}, \quad \aucNorm_i = (\numAutoi + \numOveri)(\numManui + \numOveri) - \numOveri\cdot\numOveri,
\]
where $\aucNorm_i$ is the maximum possible adjusted $\ustat$ statistic for venture $i$. When \(\numOveri = 0\), this reduces to the standard AUC: $T_i = \ustat_i/(\numAutoi \cdot \numManui)$. Finally, we combine ventures via an inverse-variance weighted mean:

\[T = \frac{\sum_{i=1}^{\numVen} T_i \cdot 1/\sigma_i^2}{\sum_{i=1}^{\numVen} 1/\sigma_i^2}
\]
where $\sigma_i^2$ is the empirical inverse variances of all scores for venture \(i\). 
This weighting ensures that ventures with more stable estimates (smaller variance) contribute more to the overall statistic.

\subsection{Two-Tailed Permutation Test}\label{section:permu}
To evaluate the statistical significance of the overall weighted AUC statistic \(T\), we employ a two-tailed permutation test with \(N = 5000\) resamples. For each venture \(i\), we keep the overlapping assignments fixed and randomly permute the labels of the manual-only and algorithmic-only assignments. This simulates the null hypothesis that assignment method has no effect on match quality.

For each permutation \(j \in \{1, \dots, N\}\), we compute the statistic \(T^j\). The two-sided p-value, $p$, is:

\[
\pval = \frac{1 + \sum_{j=1}^N \mathbb{I}\!\left[ \,\left|T^j - 0.5\right| \geq \left|T - 0.5\right| \right]}{1 + N},
\] 
which represents the proportion of permutations in which the deviation from \(0.5\) is at least as extreme as the observed deviation. An AUC value of \(0.5\) indicates that the assignment method is equally likely to be manual or algorithmic.

\end{appendix}

\end{document}